\documentclass[twocolumn,twoside, epsfig]{article}
\usepackage{graphics}
\newcommand{\hb}{\\ \hspace*{2ex}}

\begin{document}
\title{GENETIC ALGORITM ECLIPSE MAPPING}
\author{A.V.\,Halevin$^{1}$\\[2mm] %English only
\begin{tabular}{l}
 $^1$ Department of Astronomy, Odessa National University\hb
 T.G.Shevchenko Park, Odessa 65014 Ukraine, {\em halevin@odessa-astronomy.org}\\[2mm]
\end{tabular}
% no tabular, if one affiliation only
}
\date{}
\maketitle
% If the paper is submitted to another journal, but You plan to
% publish an (extended) abstract, please indicate something like
% "The complete paper is to be published in ..."
% The total width of the page is 170 mm=6.69inches= 2000 pixels (300dpi)
% The width of one column is 83 mm=3.26in=980 pixels (300dpi)
% The maximum height is 240mm=9.44in=2834pixels (300dpi)
% Please make tables and figures either for one (\begin{figure})
% or two (\begin{figure*}) dimension.
% Extensive tables may be printed with a \small font

ABSTRACT.
In this paper we analyse capabilities of eclipse mapping technique, based on genetic algorithm optimization.
To model of accretion disk we used the ``fire-flies'' conception. This model allows us to reconstruct the distribution of  
radiating medium in the disk using less number of free parameters than in other methods. Test models show that we can achieve good 
approximation without optimizing techniques.\\[1mm]
% Key words: Ierarchical structure according to the AAA list,
% the transition is marked by ":"
% different branches are separated by ";"
% same-level items are separated by ","
\\[1mm]

{\bf 1. Introduction}\\[1mm]

Eclipse mapping (EM) techniques are used for reconstruction of accretion disks structure in cataclysmic variables.
Eclipse mapping was developed by Horne (1985). One of the most important possibilities of EM is reconstruction of 
the radial temperature distribution inside accretion disk. It allows to test the different accretion disk models.
EM has now many modifications, such as flickering mapping (Bortoletto \& Baptista 2004), 3D eclipse mapping (Rutten 1998),
stream eclipse mapping with ``fire-flies'' (Hakala 2002), genetic algorithm EM (Bobinger 1999) and Physical Parameters
Mapping by Vrielmann (1997).

\begin{figure}[b]
%\resizebox{8.26cm}{!}
\resizebox{\hsize}{!}
{\includegraphics{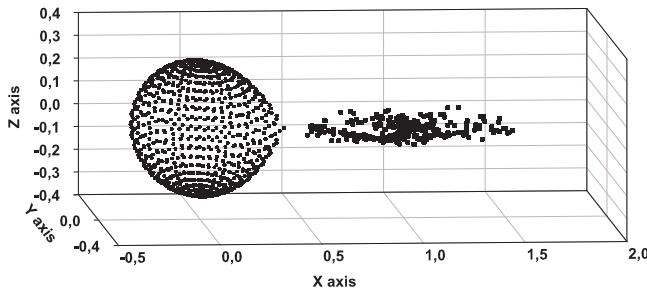}}
\label{f1}
\caption{Model scheme.}\end{figure}

In this paper we propose extension of the EM with fire-flies for reconstruction of radiating medium in the systems with flat
and optically thin accretion disks. \\[2mm]

{\bf 2. The fire-flies conception and the algorithm.}\\[1mm]

The idea of fire-flies mapping was developed by Hakala (2002) to reconstruct the structure of accretion flows in 
eclipsing polars. In this method radiation is modeled by a set of points with angle-dependent emission. Using genetic 
algorithm techniques it is possible to evolve of fire-flies spatial distribution to fit eclipse light curve by summing the
brightness of the fire-flies visible at each phase. The distribution of fire-flies gives us an emission volume. The mostly 
luminous parts of accretion stream are visible as a larger number of fire-flies placed in a smaller volume.

\begin{figure}[b]
%\resizebox{8.26cm}{!}
\resizebox{\hsize}{!}
{\includegraphics{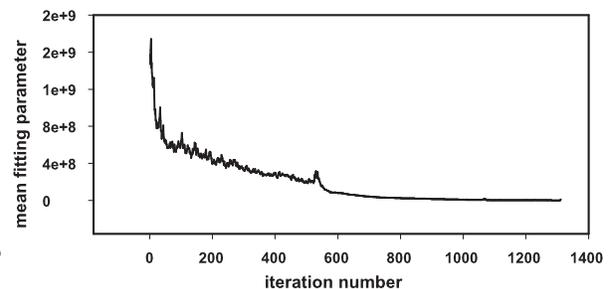}}
\label{f2}
\caption{Mean fitting parameter evolution.}\end{figure}

To model of accretion disk we used the fire-flies with isotropic emission, which reduce the number of variables, associated 
with single fire-fly to only two plane coordinates. So, in the frame of this conception, we can calculate of emission
of accretion disk with formula

\begin{equation}
   F_{disk}=\frac{F_0}{n_p}\sum_{j=1}^{n_p} E(\phi),
\end{equation}

\noindent Here $F_0$ is uncovered total accretion disk flux, $n_p$ is a number of fire-flies, $E(\phi)$ is an eclipse function,
which equal to 0 if fire-fly is eclipsed and equal to 1 if fire-fly is visible (Fig.1). 

As a fitting-function we used $\chi^2$ model parameter with some extension:

\begin{equation}
   f_i=\frac{1}{\chi^2_i+\lambda e_i}
\end{equation}

\noindent where $e_i$ is an entropy parameter, which we calculate as a mean value of first 100 minimal 
distances between pairs of particles. $\lambda$ is some constant which controls the 
importance of the smoothness for the quality determination.

\begin{figure}[t]
%\resizebox{8.26cm}{!}
\resizebox{\hsize}{!}
{\includegraphics{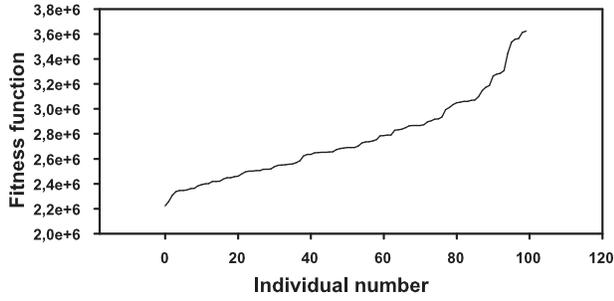}}
\label{f3}
\caption{Distribution of individuals in offspring by the fitting parameter.}\end{figure}

\begin{figure}[t]
%\resizebox{8.26cm}{!}
\resizebox{\hsize}{!}
{\includegraphics{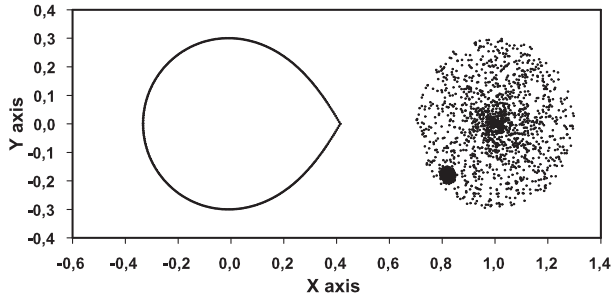}}
\label{f4}
\caption{Initial model of accretion disk with hot spot.}\end{figure}

\begin{figure}[h]
%\resizebox{8.26cm}{!}
\resizebox{\hsize}{!}
{\includegraphics{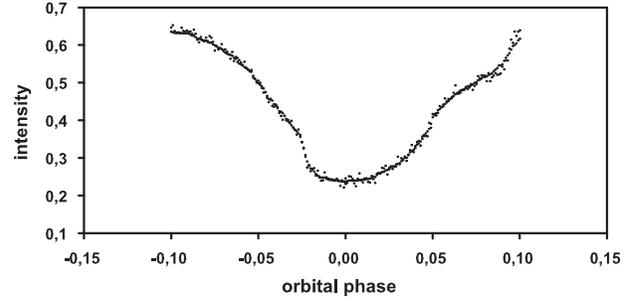}}
\label{f5}
\caption{Artificial light curve and fit for hot spot model.}\end{figure}

As a fitting algorithm we have used a Genetic algorithm (Charbonneau 1995) with 100 genes, crossover operations and variable 
mutation rate. Also we used such additional modes as conservation of the fittest individuals (so-called elitism) and 
catastrophic and Black Sheep regimes (Bobinger 2000). It takes about 1000$\div$2000 generations to achieve a good result (Fig.2). 
During calculations we control the fitting distribution of different individuals (Fig.3), to avoid degeneracy in the offspring.

\begin{figure}[t]
\resizebox{\hsize}{!}
{\includegraphics{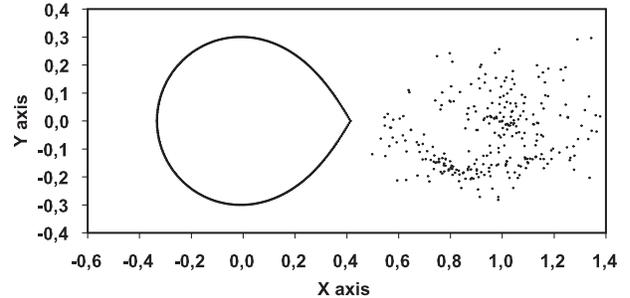}}
\label{f6}
\caption{Distribution of fire-flies for hot spot model.}\end{figure}

As a free parameters in our models are used the total disk flux $F_0$, constant non-eclipsed 
component and $n \times 2$ parameters of fire-flies coordinates. As usually, in our models we have used $n = 300\div 400$ 
to achieve reasonable computational time. Each unknown variable has 6-digits precision. Initial population is generated as 
a set of points, randomly distributed inside Roche lobe of the compact star. Actually, we have two heterogeneous sets of parameters:
main constants which describe system luminosity and coordinates of the points. To achieve a faster convergence,
during first iterations we used increased mutation rate for the first set of parameters.\\[2mm]

{\bf 3. Test models}\\[1mm]

To test of our method we have made several artificial configurations of accretion disks. The most typical situation for 
low luminosity states is the presence of three main radiation sources: the accretion disk, hot spot where accretion stream
couples with disk and filled its Roche lobe secondary star (Fig.4). In the Fig.5 one can see resulted light curve with added noise and the 
best fit (Fig.6) represents eclipse map, which have found with eclipse mapping method. One can see that main radiation sources are traced very well. 
Distortions of the image are typical for eclipse mapping techniques (see Horne 1985). So this method allows resolve 
compact sources in accretion disk.

Several cataclysmic variables show significant asymmetry of their disks which arise due to tidal effects. This feature is shown as so called 
superhumps on light curves. To test such situation we modeled elliptical accretion disk with radiation concentrated near
white dwarf (Fig.7). Eclipsing profile and model fit for such configuration one can see in the Fig.8. Resulted distribution of 
fire-flies in the Fig.9 shows very close to initial distribution picture. There is also some distortion along x-axis.

\begin{figure}[t]
%\resizebox{8.26cm}{!}
\resizebox{\hsize}{!}
{\includegraphics{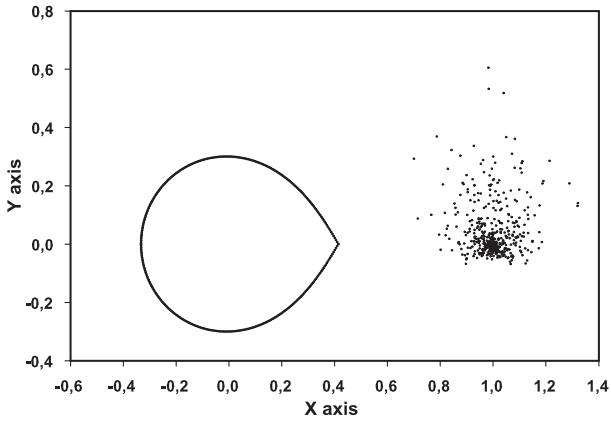}}
\label{f7}
\caption{Model of elliptical accretion disk.}\end{figure}

\begin{figure}[t]
%\resizebox{8.26cm}{!}
\resizebox{\hsize}{!}
{\includegraphics{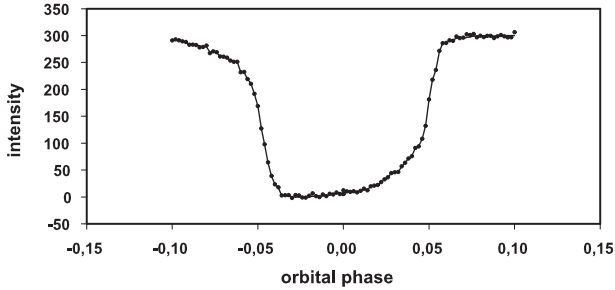}}
\label{f8}
\caption{Artificial  light curve and fit for elliptical accretion disk model.}\end{figure}

\begin{figure}[t]
%\resizebox{8.26cm}{!}
\resizebox{\hsize}{!}
{\includegraphics{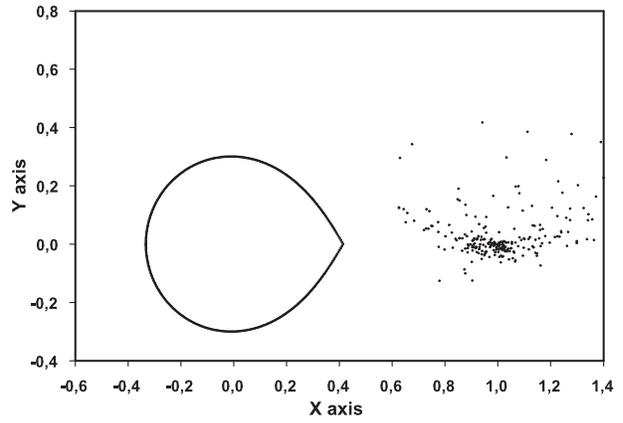}}
\label{f9}
\caption{Reconstructed fire-flies distribution for elliptical accretion disk model.}\end{figure}

Sometimes in accretion disk we can observe spiral arms, which are also product of tidal forces. This feature 
is traced only by emission lines. We tried to test reconstruction of such configuration using our modification of eclipse 
mapping. Light curve with fit and initial emission distribution are in Fig.10 and Fig.11. Fig.12 shows that we can determine 
the presence of this feature.\\[2mm]

\begin{figure}[t]
%\resizebox{8.26cm}{!}
\resizebox{\hsize}{!}
{\includegraphics{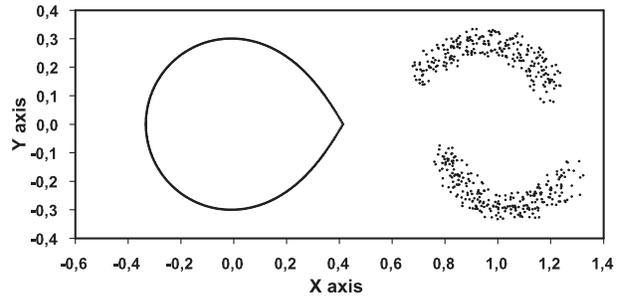}}
\label{f10}
\caption{Model of spiral arms structure in accretion disk.}\end{figure}

{\bf 4. Discussion}\\[1mm]

Here we see that the fire-flies based eclipse mapping allows reconstruct of accretion disk structure with some x-axis 
distortions, which are typical for this class of methods. It is interesting that we do not used any regularization
techniques to achieve appropriate result. We must say that our method has good convergence because several consecutive
calculations for the same light curve gave the same results. The only one important restriction for such approach 
is demand to accretion disk to be optically thin. Significant opacity in accretion disk is observed as non-eclipse
mid-time scale variability because short time scale variability, so called flickering, is the consequence of unstable processes 
in active regions of accretion disk. Previous investigators in their works used de-trending of eclipsing part of light curve 
to avoid an influence of covering effects in the disk. So, we can reconstruct only visible directly before and after the eclipse
parts of accretion disk. It is obvious that we can see source which are mainly close to the secondary star due to 
opacity effect.

So the most adequate model could be the combination of radiating fire-flies with some opaque medium. The main problem is 
how we must model of this opaque medium. If we use it distribution as a free parameter the model will be poorly conditioned.
If we use some predetermined configuration of an opaque medium it dramatically simplifies the model. In our next
papers we are going to use some models for opaque medium to fit also non-eclipsing parts of light curves.\\[2mm]

\begin{figure}[t]
%\resizebox{8.26cm}{!}
\resizebox{\hsize}{!}
{\includegraphics{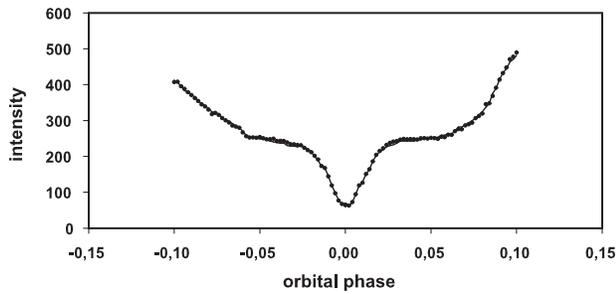}}
\label{f11}
\caption{Artificial light curve and fit for model of spiral arms in accretion disk.}\end{figure}

\begin{figure}[t]
%\resizebox{8.26cm}{!}
\resizebox{\hsize}{!}
{\includegraphics{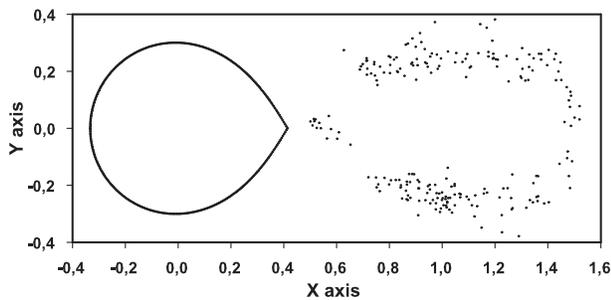}}
\label{f12}
\caption{Reconstruction of spiral arms in accretion disk.}\end{figure}

{\bf 5. Conclusions}\\[1mm]

  Using several artificial configurations typical for accretion disks of cataclysmic variables we tested fire-flies
conception based eclipse mapping technique. Our results show that we can use it successfully to reconstruct of
radiating medium distribution in optically thin flat accretion disks.
\\[2mm]

{\it Acknowledgements.} The author is thankful to S.V.Kolesnikov for helpful discussions during development
of this method.
\\[3mm]
\indent

\pagebreak

{\bf References\\[2mm]}
Bobinger A.: 2000, {\it A\&A}, {\bf 357}, 1170.\\
Bortoletto A., Baptista R.: 2004, {\it RevMexAA Conf.Ser.}, {\bf 20}, 247.\\
Charbonneau P.: 1995, {\it Ap.J. Suppl. ser.}, {\bf 101}, 309.\\
Hakala P., Cropper M., Ramsay G.: 2002, {\it A\&A}, {\bf 334}, 990.\\
Horne K.: 1985, {\it MNRAS}, {\bf 213}, 129.\\
Horne K.: Stiening R., 1985, {\it MNRAS}, {\bf 216}, 933.\\
Rutten R.G.M.: 1998, {\it A\&ASupl}, {\bf 127}, 581.\\
Vrielmann S.: 1997, {\it PhD Theses}.\\

%\vfill
%the final page should be formatted to make the height of the column
%nearly equal. This may be done by splitting the line in the first
%column e.g. by a command
%\linebreak\vfill\pagebreak\noindent
\end{document}